\documentclass[11pt,letterpaper]{article}
\usepackage[margin=1in]{geometry}
\usepackage{color}
\usepackage{enumitem}
\usepackage{titling}
\usepackage{comment}
\usepackage{tabularx}
\usepackage{booktabs}
\usepackage{caption}
\usepackage[color=gray!10, size=tiny, textsize=tiny]{todonotes}
\usepackage{hyperref}
\usepackage{wrapfig}
\usepackage[linesnumbered,lined,ruled,noend]{algorithm2e}
\usepackage{amsmath,amssymb}
\usepackage{stmaryrd}
\usepackage{graphicx}
\usepackage{subcaption}
\usepackage{cleveref}

\usepackage{xspace}
\usepackage{mathtools}
\usepackage{amsthm}

\newtheorem{theorem}{Theorem}
\newtheorem{lemma}{Lemma}

\newcommand{\Math}[1]{\ensuremath{#1}\xspace}
\newcommand{\E}{\Math{\mathcal{E}}}

\newcommand{\Enc}{\mathsf{Enc}}

\newcommand{\en}[1]{\Math{\llbracket #1\rrbracket}}

\newcommand{\hashTable}{{H}}
\newcommand{\compare}[2]{\mathsf{compare}\left(#1, #2\right)}

\newcommand{\circler}{\textsuperscript{\textregistered}}

\newcommand{\Tdb}{D^S} 
\newcommand{\Adb}[1]{D^{B_{#1}}} 

\newcommand{\TX}{\Math{\mathcal{X}}} 
\newcommand{\tx}{\Math{\mathsf{tx}}} 
\newcommand{\FS}{\Math{F_S}} 
\newcommand{\FB}{\Math{F_{B}}} 
\newcommand{\fj}[1]{\Math{f_{#1}}} 
\newcommand{\Acck}[1]{\Math{a_{#1}}} 
\newcommand{\flag}{\Math{\mathsf{flag}}} 
\newcommand{\Acc}{\Math{\mathcal{A}}} 
\newcommand{\PET}[2]{\mathsf{PET}\left(#1, #2\right)} 

\newcommand{\sk}{\mathsf{sk}} 
\newcommand{\pk}{\mathsf{pk}} 

\newcommand{\myparagraph}[1]{\medskip\noindent\textbf{#1}}

\newcommand{\NlL}[2]{\Math{N^{#1}{[#2]}}} 
\newcommand{\Label}[1]{\Math{\mathsf{label}(#1)}} 
\newcommand{\features}[1]{\Math{\mathsf{features}(#1)}} 
\newcommand{\tests}[1]{\Math{\mathsf{tests}\left(#1\right)}} 
\newcommand{\C}[2]{\Math{\mathcal{C}_{#1}(#2)}} 
\newcommand{\numBins}{\Math{\mathsf{numBins}}} 

\newcommand{\Real}[3]{\mathsf{REAL}^{#1}_{#2}{\left(#3\right)}}
\newcommand{\Sim}[3]{\mathsf{SIM}^{#1}_{#2}{\left(#3\right)}}

\newcommand{\CS}{C_S}
\newcommand{\CB}{C_B}

\title{Privacy-Preserving Federated Learning over Vertically and Horizontally Partitioned Data for Financial Anomaly Detection}
\author{Swanand Ravindra Kadhe, Heiko Ludwig, Nathalie Baracaldo, Alan King,  Yi Zhou, \\ 
Keith Houck, Ambrish Rawat,  Mark Purcell, Naoise Holohan, Mikio Takeuchi, \\
Ryo Kawahara, Nir Drucker, Hayim Shaul, Eyal Kushnir, Omri Soceanu
\\ IBM Research }

\date{\vspace{-5ex}}

\begin{document}
\maketitle


\begin{abstract}
The effective detection of evidence of financial anomalies requires collaboration among multiple entities who own a diverse set of data, such as a payment network system (PNS)  and its partner banks. Trust among these financial institutions is limited by regulation and competition. However, all are aligned in their desire to improve the detection of suspicious transaction anomalies. 
Federated learning (FL) enables entities to collaboratively train a model when data is either vertically or horizontally partitioned across the entities. However, in complex, real-world financial anomaly detection scenarios, the data is partitioned both vertically \textit{and} horizontally and, hence, it is not possible to use existing FL approaches in a plug-and-play manner. 

Our novel solution, PV4FAD, combines fully homomorphic encryption (HE), secure multi-party computation (SMPC), differential privacy (DP), and randomization techniques to balance privacy and accuracy during training and to prevent inference threats at model deployment time. 
Our solution provides \textit{input privacy} through HE and SMPC, and \textit{output privacy} against inference time attacks through DP. 
Specifically, we show that, in the \textit{honest-but-curious} threat model, banks do not learn any sensitive features about PNS transactions, and the PNS does not learn any information about the banks' dataset but only learns prediction labels. 
We also develop and analyze a DP mechanism to protect output privacy during inference.
Our solution generates high-utility models by significantly reducing the per-bank noise level while satisfying distributed DP.
To ensure high accuracy, our approach produces an \textit{ensemble model}, in particular, a \textit{random forest}. This enables us to take advantage of the well-known properties of ensembles to reduce variance and increase accuracy. 
Our solution mitigates potential loss in accuracy with DP techniques by taking advantage of random sampling and boosting techniques that carefully select subsets of data samples to train a model. 
Furthermore, the ensemble model can be built considering different and complex phenomena required for financial crimes detection. 
Finally, the entire stack is designed to be efficient and scalable.

Our solution won second prize in the first phase of the U.S. Privacy Enhancing Technologies (PETs) Prize Challenge. 
 

\end{abstract}

\section{Introduction}
\label{sec:intro}

Money laundering and other financial crimes, add up to a roughly \$2 trillion hit to the global economy every year~\cite{weforum, unitednations}. Only a small fraction of these illegal transactions are actually detected as such, which limits the value of illicit assets that end up being frozen by law enforcement authorities to about 1\%.

One way to improve the detection of suspicious transactions is through more systematic data sharing among banks and other institutions within the international financial system, such as a payment network system. Applying machine learning and AI techniques on shared data can provide a great deal of insight towards tracking the money trail of criminal activity of all sorts. However, there are good reasons for financial organizations to be reluctant about openly sharing transaction data with one another. Some of these concerns are regulatory, as well as concerns about losing competitive advantage. 
Such conflicting requirements create a tension between enabling sufficient access to machine learning techniques to effectively detect anomalous financial activity, and limiting the leakage of sensitive information that may result from data sharing.

Federated Learning (FL) has recently emerged as a distributed learning paradigm that enables multiple parties to coordinate with an aggregator (central server) to train a shared model while keeping all the training data on premise. FL provides some degree of privacy by allowing each party to keep their training data on premise and share only model updates or gradients. However, the level of privacy provided by conventional FL techniques is limited. In particular, it has been shown that model updates or gradients from the parties may still leak substantial information making them vulnerable to inference and inversion attacks~\cite{fredrikson2015model, shokri2017membership, nasr2019comprehensive, ganju2018property}. Privacy guaranties can be significantly enhanced by integrating FL with either Secure Multiparty Computation (SMPC) techniques, Differential Privacy (DP) techniques or a combination thereof~\cite{ludwig2022federated, kairouz2021advances, li2020federated}. However, nearly all existing privacy-enhancing approaches for FL assume that the data is either horizontally partitioned (i.e., parties have different samples) or vertically partitioned (i.e., parties have different features of the same set of samples). This assumption does not hold for complex, real-world scenarios such as financial anomaly detection. 

\begin{figure}[ht!]
    \centering
    \begin{subfigure}{0.9\textwidth}
    \centering
    \includegraphics[scale=0.13]{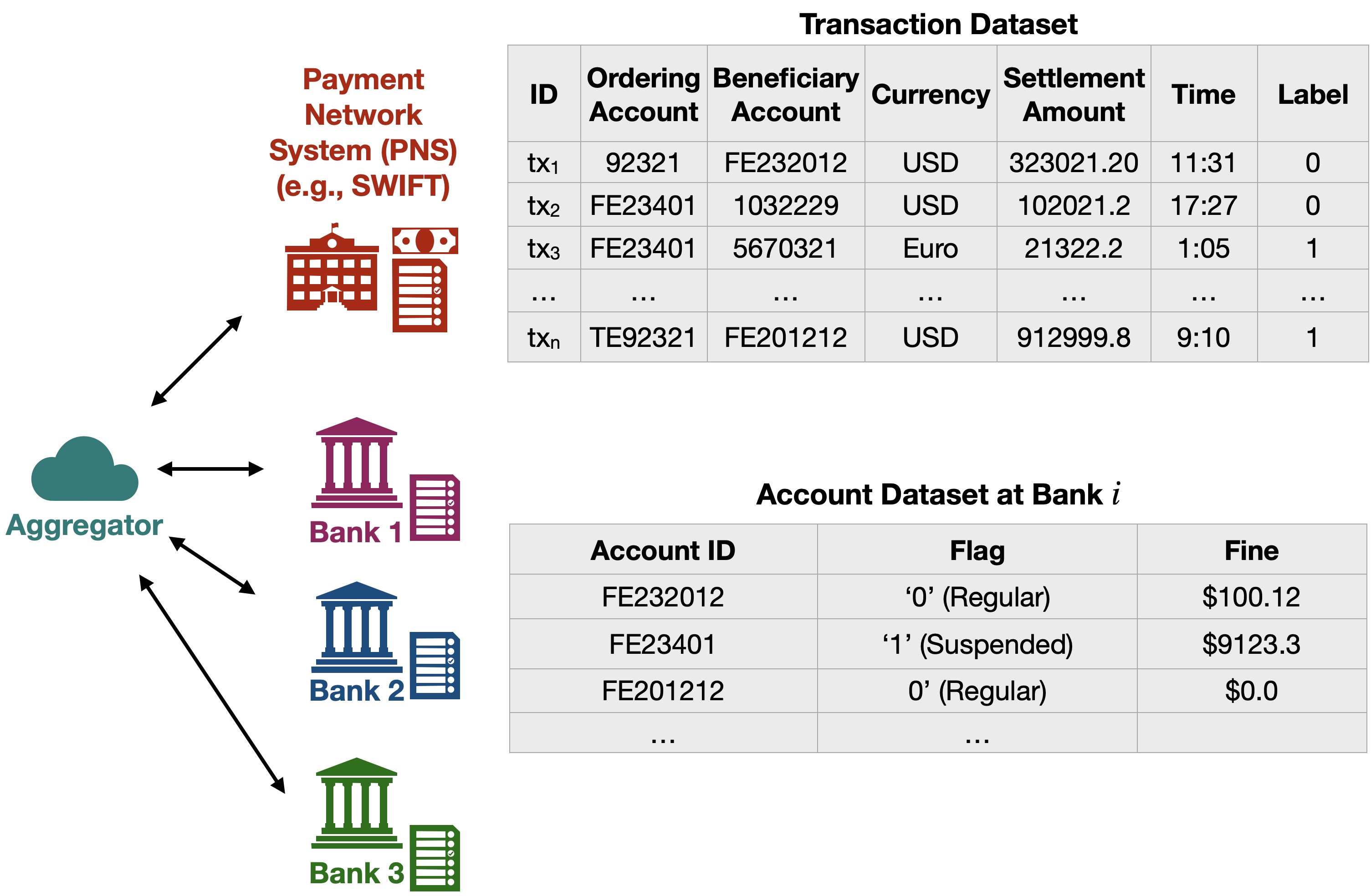}
    \caption{Federated Learning for financial anomaly detection, involving a Payment Network System (PNS), and its partner banks. The PNS has labeled transactions while the banks have account information.}
    \label{fig:FL-scenario}
    \end{subfigure}
    \\
    \begin{subfigure}{0.9\textwidth}
    \centering
       \includegraphics[scale=0.13]{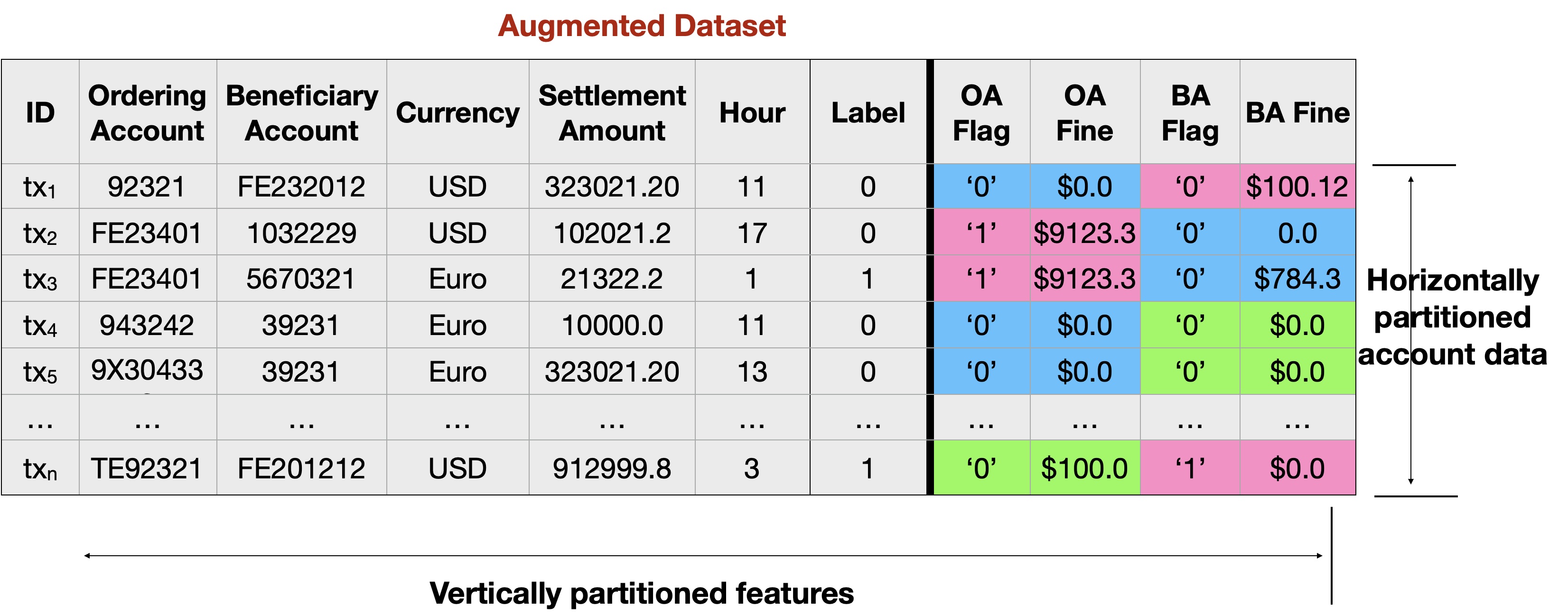}
    \caption{Combined data from the PNS and banks is vertically partitioned -- account features are with the banks and transactions features and labels are with the PNS, as well as horizontally partitioned – accounts are partitioned across banks. Here, OA denotes Ordering Account, BA denotes Beneficiary Account, and colors are used to indicate banks that hold corresponding accounts (e.g., account IDs starting with `FE' are with Bank 1 represented using magenta color).}
    \label{fig:data-partitioning}
    \end{subfigure}
    \caption{Privacy-Preserving Federated Learning on Vertically and Horizontally Partitioned Data}
    \label{fig:FL-scenario-and-data-partitioning}
\end{figure}

Financial anomaly detection requires privacy-preserving federated learning solutions when the data is \textit{both vertically and horizontally partitioned} across multiple financial institutions during training as well as inference. Consider a typical scenario wherein a Payment Network System (PNS) wants to collaborate with its partner banks to train a model for financial anomaly detection (Fig.~\ref{fig:FL-scenario}). The PNS has access to transactions along with some transactions labeled with anomalies. Each transaction typically has several attributes including \textit{Ordering Account}, \textit{Beneficiary Account}, amount, currency, etc. Banks have access to account information, which consists of account identifiers and attributes such as flags (indicating whether the account is valid, suspended, etc.) and fine amounts. From the perspective of federated learning, the data are vertically partitioned -- the features of \textit{Ordering} and \textit{Beneficiary} accounts are with the banks, as well as horizontally partitioned – accounts are partitioned across banks (Fig.~\ref{fig:data-partitioning}). 

The problem of financial anomaly detection poses several unique \textbf{challenges}: First, 
the mix of vertical and horizontal data partitioning makes it ill-suited for existing privacy-enhancing approaches for vertical FL or horizontal FL, e.g., \cite{xu2021fedv, cheng2021secureboost, hardy2017private, truex2019_hybrid, bonawits2017_practical, so2020turbo, kadhe2020_fastsecagg}, in a plug-and-play manner. The data do not follow traditional entity resolution or private set intersection assumptions \cite{hardy2017private, yang2019federated, pprl2022}. Multiple transactions may link to the same account; hence, creating appropriate feature vectors requires \emph{replicating} records at the bank side as opposed to the usual aligning and reordering processes of standard VFL, more akin to a database join. This poses challenges that, to the best of our knowledge, have not been addressed efficiently by existing state of the art software-based approaches \cite{ludwig2020ibm, liu2021fate,romanini2021pyvertical,he2020fedml}.

Second, training highly accurate models for detecting transaction anomalies poses challenges on the machine learning side, which directly impacts the design of privacy solutions. Since anomalous transactions are often much fewer than benign ones, training data comes with a low signal-to-noise and a high class imbalance. Therefore, straightforwardly using the attributes of  transactions as features can be significantly limiting in terms of the performance of the trained model, and it is important to derive augmented features. For example, organizing transaction data into a graph, where nodes represent accounts (transaction end points) and edges represent the transactions between accounts, and deriving statistical features of graph nodes and edges may boost the model performance~\cite{suzumura2022federated,molloy2016graph}. Thus, it is crucial to design privacy solutions that can enable training models on such complex engineered features without compromising the privacy of  highly sensitive features. Furthermore, the choice of ML models and whether any approximations are required by the privacy solutions used play a vital role in governing the utility of trained models.

\textbf{Our Contributions:} We develop a novel privacy-preserving federated learning solution for financial anomaly detection that addresses the aforementioned challenges. 
Our holistic approach combines fully homomorphic encryption, secure multi-party computation (SMPC), differential privacy (DP), and randomization techniques to balance privacy and accuracy during training and to prevent inference threats at model deployment time. To ensure high accuracy, our approach produces an ensemble model, in particular, a random forest. We propose to utilize a random forest consisting of \textit{random decision trees}, motivated by their favorable privacy versus utility trade-off. 

Our solution provides \textit{input privacy} through homomorphic encryption and \textit{output privacy} against inference time attacks through differential privacy. Specifically, we provide privacy proofs in the honest-but-curious threat model to show that: the banks do not learn any sensitive features about the PNS transactions; the PNS does not learn anything about the account dataset at any bank, it only learns prediction labels at  inference time; and the aggregator does not learn anything about  transactions at the PNS as well as account datasets at the banks. One of the key novel components of our solution is a Private Intersection Sum (PI-Sum) protocol based on homomorphic encryption, which may be of independent interest.

To protect \textit{output privacy} during inference, we also develop and analyze a differential privacy (DP) mechanism that allows each bank to add calibrated Laplace noise. Our solution generates high-utility models by significantly reducing the per-bank noise level while satisfying distributed DP.

We submitted this solution to the U.S. Privacy Enhancing Technologies (PETs) Prize Challenge\footnote{\url{https://www.drivendata.org/competitions/group/nist-federated-learning/}} (Financial Crime Track), and it won second place among the winners of the challenge's first phase\footnote{\url{https://drivendata.co/blog/federated-learning-pets-prize-winners-phase1}}. 

We note that, while we focus our attention on the financial anomaly detection scenario, our solution and its component techniques can be applied, in principle, to any scenario for performing privacy-preserving federated learning over vertically and horizontally partitioned data. For instance, our solution can enable an insurance provider company and its partner hospitals, clinics, and test laboratories to collaboratively train a model for accepting/rejecting claim applications, without compromising the privacy requirements. 

\section{Threat Model and Privacy Objectives}
\label{sec:threat-model}


\noindent Our \textbf{threat model} includes the following entities: PNS, banks, and an aggregator. We focus on the \textit{honest-but-curious} threat model with computationally bounded adversaries, where each participant honestly follows the protocol, but attempts to learn about the sensitive dataset belonging to the other entities by using the messages exchanged during the execution of the protocol. The honest-but-curious model is commonly used in prior work on secure federated learning~\cite{bonawits2017_practical, truex2019_hybrid, kadhe2020_fastsecagg}. 

We assume that the transaction and account data are not maliciously crafted and any errors are non-malicious; poisoning attacks are outside of the scope of our solution.
During the pre-processing phase, wherein PNS and banks may extract and engineer features, we assume that all entities generate genuine features. 
We do not consider any privacy issues at that phase, and assume that integrity is promised by treating the entities as semi-honest (or covert) adversaries that do not have an incentive to provide erroneous data to the latter phases of the protocol. 
For simplicity of exposition, we assume that any incomplete or inconsistent entries in the datasets are removed during pre-processing.  


\vspace{4pt}
\noindent Our goal is to develop a  solution guided by the following \textbf{objectives}:

\vspace{2pt}
\noindent\textbf{1. Strong privacy guarantees during training:}
Consider a hypothetical scenario wherein PNS and banks can take help of a trusted third party (TTP) that provides ML-as-a-service.  PNS and banks send their data to TTP, which trains a model and keeps the model with it.  During inference, PNS sends transactions to TTP, which generates labels using the model and shares them with PNS. Our goal is to design a private protocol that can emulate  a TTP \textit{without a TTP}.

\vspace{2pt}
\noindent\textbf{2. Privacy during inference without downgrading accuracy:} Even if an algorithm provides privacy during training, PNS can potentially learn information about bank datasets during inference and vice-versa. Recent work has shown that membership inference attacks can be launched even if the adversary uniquely has access to the labels of the prediction~\cite{choquette-choo2021_label, zheng2021_membership}. A principled method to defend against inference-time attacks is to train a model with differential privacy (DP).  We want to judiciously select model architectures and carefully design privacy protocols to achieve DP without compromising accuracy.

\vspace{2pt}
\noindent\textbf{3. Ability to privately train models on complex features:} As discussed in Sec.~\ref{sec:intro}, incorporating complex graph features engineered from transaction data can play a vital role in boosting the model performance. Thus, we want to design a protocol that can incorporate such complex features and train models while preserving the privacy of these highly sensitive features.

\section{Problem Setup} 
\label{sec:problem-setup}

Our setup consists of a Payment Network System (PNS) $S$, along with $N$ banks $B_1, B_2, \ldots, B_N$, and an aggregator.
The goal is to develop a federated learning solution that allows the PNS and banks to coordinate with the aggregator to train a model to classify \textit{anomalous transactions}. Anomalous transactions are essentially \textit{outlier} transactions that significantly differ from typical transactions, and thus, such anomalous transactions may be indicative of a financial crime such as money laundering or fraud. For example, transactions that include unexpected currency or amount, atypical timestamps, inconsistent account attributes, or unusual senders/receivers (i.e., corridors). We assume that the PNS has a transaction dataset labeled with anomalies, which is used during training.

In particular, the PNS has a dataset $\Tdb$ of transactions, where each transaction has several attributes including \textit{Ordering Account}, \textit{Beneficiary Account}, transaction ID, amount, currency, time, etc. Each bank $B_i$ has its account dataset $\Adb{i}$, where each row corresponds to an account identifier and columns include attributes such as flags and fine amounts. The datasets can be linked using the account identifiers in the bank data and the fields---Ordering Account and Beneficiary Account---in the PNS transaction data (see, e.g., Fig.~\ref{fig:FL-scenario-and-data-partitioning}). 

For concreteness, we describe our solution by focusing on the data format used in the Financial Crime Track of the PETs Prize Challenge\footnote{\url{https://www.drivendata.org/competitions/105/nist-federated-learning-2-financial-crime-federated/page/589/}}, which used the synthetic dataset provided by SWIFT\footnote{\url{https://www.swift.com/}}.




\section{Our Solution: Private Vertical Ensemble}
\label{sec:technical-approach}

\begin{figure}[h]
\centering
\includegraphics[width=0.97\textwidth]{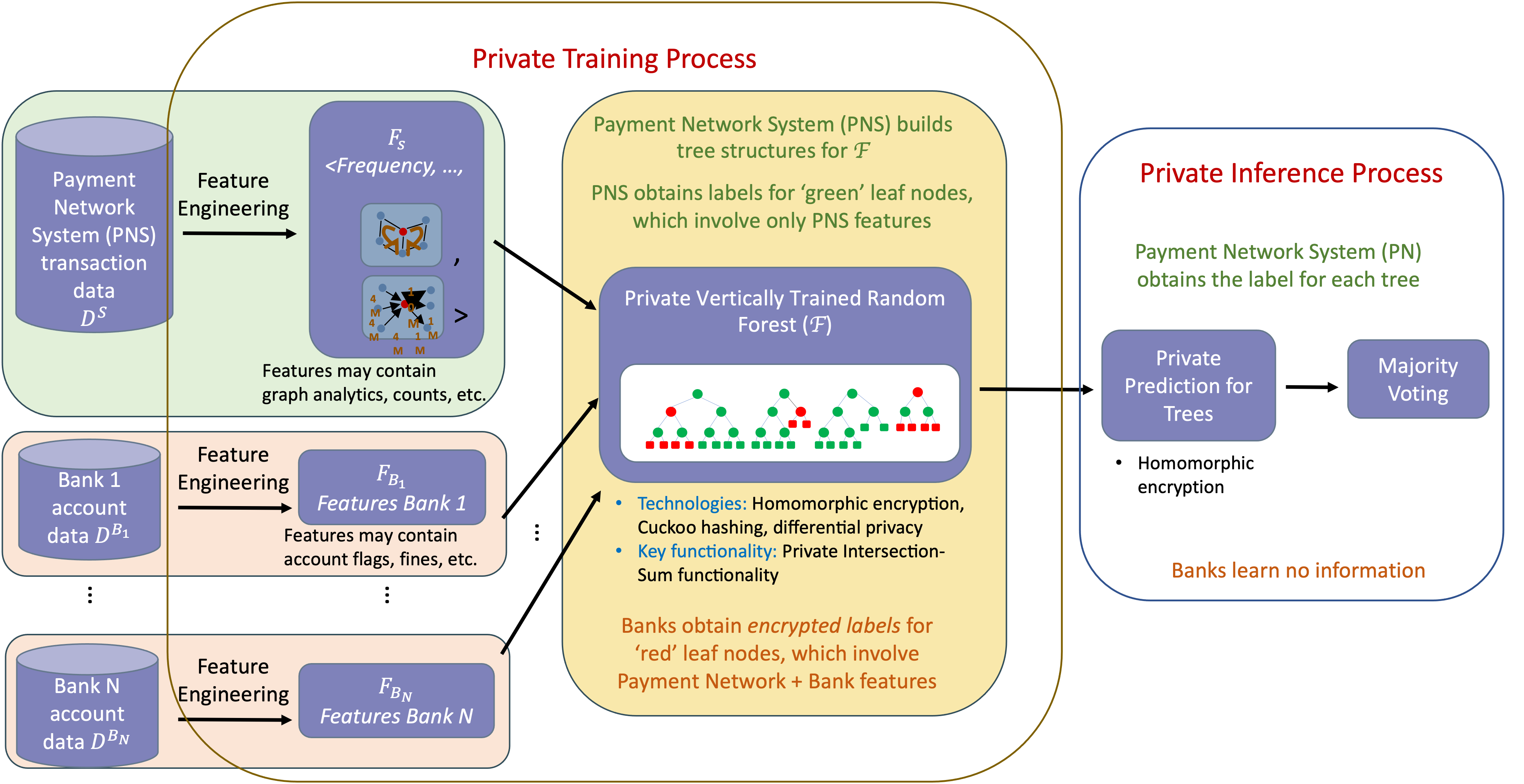}
\caption{Overview of the system consisting of a private training and a private inference processes.} \label{fig:overview}
\end{figure}

Our solution PV4FAD (Private Vertical Ensemble for Financial Anomaly Detection) is a holistic approach that combines several cryptographic, privacy, and machine learning techniques to generate a \textit{random forest} with the help of an Aggregator. Fig. \ref{fig:overview} provides an overview of the system. We begin with outlining the key components and then present the detailed protocol.

\vspace{4pt}
\noindent\textbf{Model Architecture:} Our privacy solution enables PNS and banks to collaboratively train an ensemble model, in particular a random forest, without learning anything about each others' private datasets. Choosing an ensemble model enables us to take advantage of the well-known properties of ensembles to reduce variance and increase accuracy.
Conventionally, a random forest consists of \textit{greedy} decision trees, where features in a tree are chosen greedily using some judiciously defined criterion, e.g., information gain~\cite{shalev-shwartz2014_understanding}. We propose to train a random forest consisting of  \textit{random decision trees}~\cite{Fan2003_is,Geurts2006_extremely}. 
In a random decision tree, features for the tree nodes are chosen at random instead of using a selection criterion. The structure of a random decision tree is built independently of the training data. The training data is used only to determine labels associated with the leaf nodes of the tree.

\textit{Our choice of utilizing random decision trees (RDTs) is motivated by their favorable privacy versus utility trade off.} First, we note that random forests built with RDTs have been shown to be competitive against conventional random forests even in non-private scenarios~\cite{Fan2003_is,Geurts2006_extremely,Fletcher2019_decision}. Second, RDTs are an excellent fit with privacy enhancing technologies \cite{crf}. To highlight this, we note that, for a conventional decision tree, even the structure of the tree leaks information. For instance, suppose that a trusted third-party trains a (conventional) greedy decision tree and places a beneficiary account flag feature as the root. Then, this tree structure leaks to PNS that beneficiary account flags are highly correlated with the transaction labels. There are two ways to reduce such leakage in conventional greedy decision trees: (i) Use DP mechanisms when choosing features. However, it has been shown that this negatively impacts the accuracy~\cite{Jagannathan2012_apractical}, and RDTs with DP can achieve better prediction accuracy than their greedy counterparts~\cite{Friedman2010_data,Rana2015_differentially,Fletcher2017_differentially,Fletcher2019_decision}. (ii) Use secure MPC techniques to compute information gain. However, this greatly increases the computation complexity and limits scalability~\cite{Fletcher2019_decision}. Additionally, our solution circumvents the necessity of performing private set intersection or record alignment which is typically computationally heavy.

\vspace{4pt}
\noindent\textbf{Private Feature Engineering:} 
Our solution allows PNS and each bank to locally engineer complex features. As discussed in the privacy objectives in Sec.~\ref{sec:threat-model}, incorporating statistical features of transaction graphs, including attributes of account nodes and their neighborhoods, can significantly boost the accuracy of the trained model.
On the PNS side, we apply a pipeline of proven graph-based financial crime detection techniques
\cite{weber2018scalable} to the PNS data and feed these results into an ensemble of privacy-preserving decision trees to incorporate the influence of the bank data without exposing the latter to PNS (or vise versa). Note that the features extracted at a participant remain locally at the participant, and the training, as well as inference protocols, are designed to preserve the privacy of the features from the other participants including the aggregator.


\vspace{4pt}
\noindent\textbf{Privacy-Preserving Training:} A benefit of using RDTs is that tree structures can be built \textit{independently of the training data}.
For ease of exposition, we assume that PNS builds the tree structures (this assumption can be removed). The most challenging part of the training process is to (privately) compute the label for each leaf node, which may depend on both PNS and bank features. 
We observe that the leaf nodes of every tree can be partitioned into two types: (i) \textit{`green' leaf nodes} -- they lie on a path containing \textit{only} PNS features; and (ii) \textit{`red' leaf nodes} -- they lie on a path containing PNS \textit{and} Bank features. 
As we describe in the next section, PNS can locally compute the labels for all `green' leaf nodes. 
We develop a novel protocol based on Homomorphic Encryption that enables PNS and banks to collaborate for computing the labels of `red' leaf nodes (Sec.~\ref{sec:priv-RDT-training}). At the end of this protocol, banks obtain the encrypted label for each `red' leaf node, where the secret-key is with PNS. 
The PNS does not learn any information about any bank's account dataset, and any bank does not learn any information about PNS's transaction dataset or the other banks' account datasets. In other words, our protocol mimics a trusted third party that computes labels and keeps the labels with it. We provide rigorous privacy guarantees in Sec.~\ref{sec:privacy}.


\vspace{4pt}
\noindent\textbf{Privacy-Preserving Inference:} Since the labels of `red' leaf nodes are with banks, PNS needs to collaborate with the banks during inference. We develop a protocol based on HE that enables PNS to obtain the label of the appropriate leaf node (Sec.~\ref{sec:priv-RDT-inference}). 

\vspace{4pt}
\noindent\textbf{Protecting against Inference-Time Attacks:} To protect against inference time attacks, we incorporate differential privacy by building on the techniques in~\cite{Jagannathan2012_apractical}, wherein each bank adds calibrated Laplace noise, when computing the number of labels of `red' leaf nodes under HE (Sec.~\ref{sec:privacy}).

\subsection{Privacy-Preserving Training}
\label{sec:priv-RDT-training}

\begin{wrapfigure}[19]{r}{7.75cm}
    \centering
    \includegraphics[width=0.425\textwidth]{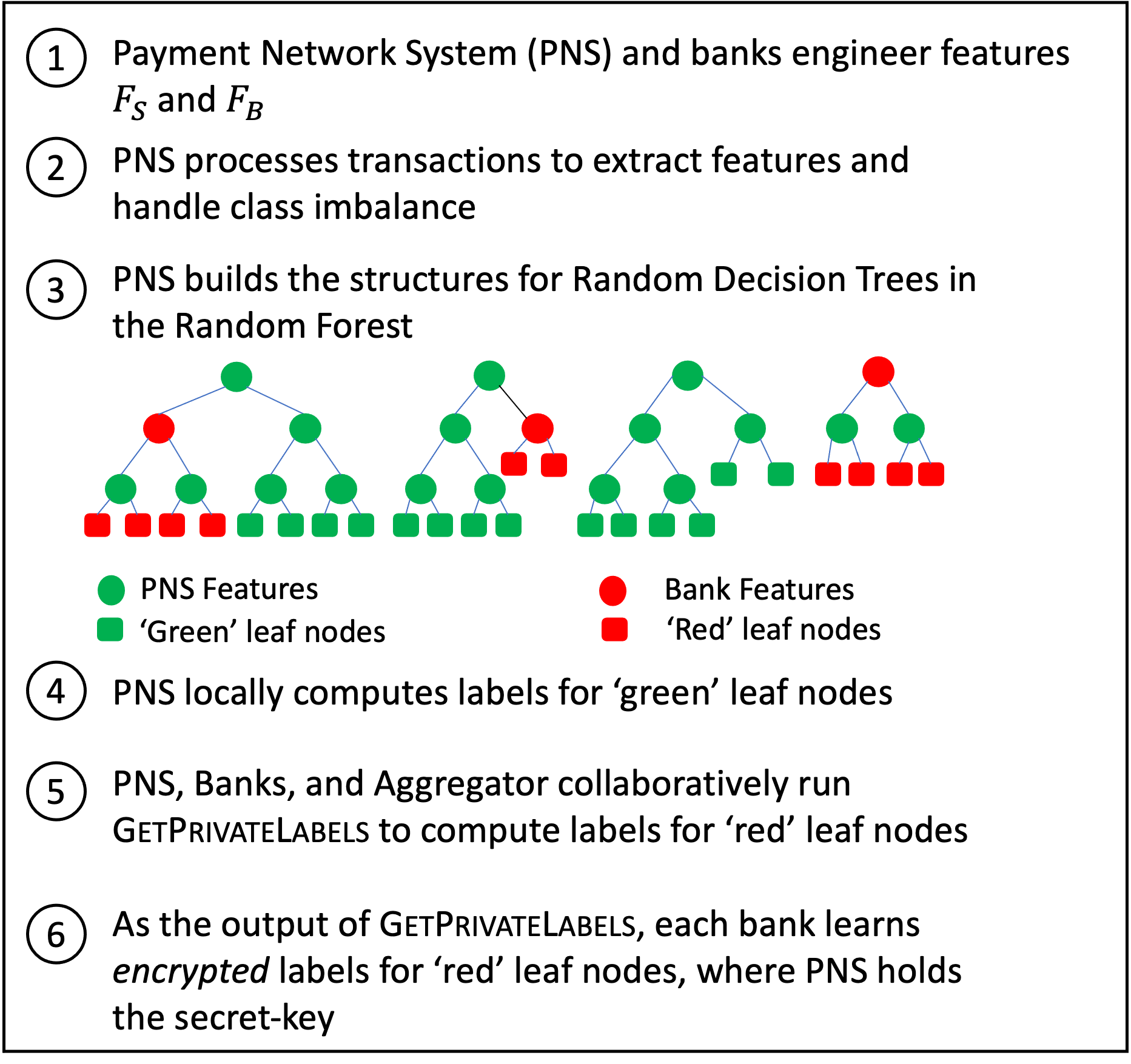}
    \caption{ \label{fig:priv-RDT-train} {Protocol for Privacy-Preserving Training of Random Forest.}}
\end{wrapfigure}

Our end-to-end protocol for privately training a random forest is described in Fig.~\ref{fig:priv-RDT-train}. 
During the pre-processing phase, PNS engineers transaction features $\FS$ with each row being a transaction entry, and processes its transactions to extract their features. PNS can perform resampling and other data augmentation techniques to handle class imbalance~\cite{japkowicz2000class,japkowicz2002class,liu2008exploratory}.
PNS also selects hyperparameters for the random forest: the number of trees and the height of each tree.

The next step is to build tree structures. For this step, we assume, for simplicity, that PNS and banks agree on common features $\FB$ across all banks. We further restrict our attention to the case when $\FB = \{$Ordering Account Flag, Beneficiary Account Flag$\}$. This assumption can be lifted to enable each bank to engineer their private features, where PNS samples `slots' for banks who place their features in these slots (we omit the details for space). Based only on $\FS\cup\FB$, PNS builds the structure of each RDT recursively, where a feature is selected randomly for each node, and a \emph{test} (a.k.a. a comparison or rule) is selected for each edge. For categorical features, an edge is added for each possible value, while for continuous features splitting points are chosen randomly (see~\cite{Fan2003_is,Geurts2006_extremely} for details). 
It is important to note that PNS builds tree structures independent of transaction and account datasets. Therefore, any tree structure does not leak any information about the training data. Moreover, tree structures are never shared with banks and thus they do not learn anything about PNS features $\FS$.  

The last training step is to compute labels for the leaf nodes of each RDT. We first clarify some notation we will use throughout the rest of this section.

For a leaf node $L$, let $\features{L}$ denote the set of features that lie on its path from the root, and $\tests{L}$ denote the set of tests associated with the edges on its path from the root. For simplicity, we assume that, for any leaf $L$, $\features{L}$ contains at most one bank feature from $\FB$. We emphasize that our protocols can be generalized to remove this assumption.

\begin{wrapfigure}[19]{r}{7cm}
    \centering
    \includegraphics[width=0.425\textwidth]{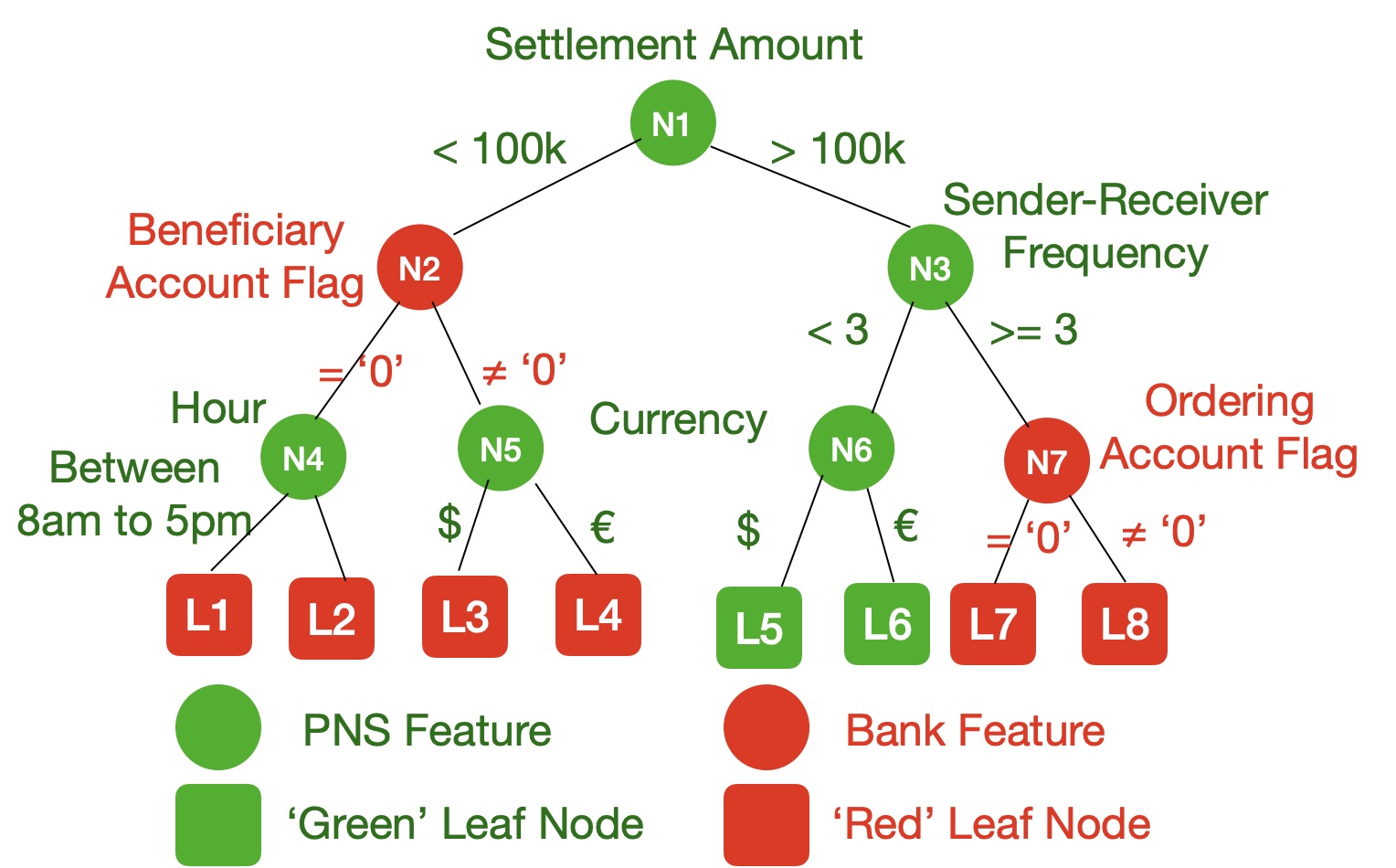}
    \caption{ \label{fig:RDT} {Toy example of a decision tree.}}
\end{wrapfigure}    

Also note that leaf nodes across all RDTs can be partitioned into two sets: (i) set $\mathcal{L}_S$ with leaf nodes depending \textit{only} on PNS features $\FS$, i.e., any leaf node $L$ such that $\features{L}\subseteq\FS$. We refer to these leaf nodes as \textit{`green' leaf nodes}; (ii) set $\mathcal{L}_B$ with leaf nodes depending on bank (and possibly PNS) features, i.e., any leaf node $L$ such that $\features{L}\cap\FB \ne \varnothing$. We call them as \textit{`red' leaf nodes}.
PNS can locally compute the label for each `green' leaf node, since it involves only PNS features. To generate the label for a leaf $L$, $\Label{L}$, PNS computes the number of label-$0$ and label-$1$ transactions $\NlL{0}{L}$ and $\NlL{1}{L}$ that reach $L$, respectively. (A transaction reaches a leaf node if it satisfies all the tests in $\tests{L}$.) If $\NlL{0}{L} \geq \NlL{1}{L}$, then $\Label{L} = 0$; else, $\Label{L} = 1$.   See Fig.~\ref{fig:RDT} for an example.

Finally, PNS, banks, and the aggregator 
run an SMPC protocol \textsc{GetPrivateLabels} (Fig.~\ref{fig:get-private-labels}) to obtain labels for the `red' leaf nodes. As the output of \textsc{GetPrivateLabels}, each bank obtains the  \textit{encrypted} label for each `red' leaf node, where the secret-key is with PNS.
The key component of obtaining the label at a leaf node in \textsc{GetPrivateLabels} is to compute the number of transactions with a given label that can reach the leaf node. We build this functionality on top of \textit{Private Intersection Sum} (PI-Sum) by designing a PI-Sum protocol based on fully homomorphic encryption. Although recent work on PI-Sum has arisen  (e.g.,~\cite{ion2017_private,pinkas2019_efficient,ion2020_on, psswc2}), these methods cannot be directly applied to our case because they necessitate both parties to learn the cardinality of the set intersection in plaintext, which can cause significant privacy leakage in our case.

\begin{figure}[!h]
\small
\caption{
	\textsc{GetPrivateLabels} for privately generating the labels of `red' leaf nodes in $\mathcal{F}$.}
	\label{fig:get-private-labels}
	\vspace{-0.4cm}
\fbox{
	\begin{minipage}{0.97\textwidth}
	\myparagraph{Participants:} PNS $S$, Banks $B_1, \ldots, B_N$, Aggregator
	
	\myparagraph{Public Parameters:} a security parameter $\lambda$, a DP parameter $\epsilon$, dataset schema, an HE scheme $\E_S$, set of $h$ hash functions $\mathsf{hash}_h:\{0,1\}^{\sigma} \rightarrow [0, \numBins]$
	
	\myparagraph{Inputs at PNS:} Random Forest $\mathcal{F}$, Set of `red' leaf nodes $\mathcal{L}_B$
	
	\myparagraph{Inputs at Banks:} Bank $B_j$ has set $\Acc_{B_j}$ of pairs of account number and its flag
	
    \myparagraph{Output:} Each bank gets encrypted labels for the `red' leaf nodes $\{\en{\Label{L}}:L\in\mathcal{L}_B\}$
    
    \myparagraph{Setup:} PNS generates a key-pair $(\sk, \pk)$ for HE $\E_S$ and shares $\pk$ with Banks and Aggregator
		
	\begin{enumerate}
	    \item PNS computes, for each `red' leaf node $L\in\mathcal{L}_B$ and label $\ell\in\{0,1\}$: 
	    \begin{enumerate}
	        \item The bank feature $\fj{b}[L]$ (e.g. Ordering Account Flag or Beneficiary Account Flag), and the feature value $\mathsf{tests}_{f_b}[L]$ (e.g., flag value ) associated with $L$.
	        
	        \item The sets of label-$\ell$ candidate transactions $\TX^{\ell}[L]$, more specifically,  $\TX^{\ell}[L]$ is the subset of transactions that satisfy all the tests in $\tests{L}$ except the test associated with $\fj{b}$.
	        
	        \item The set $\Acc^{\ell}[L] = \{(\Acck{i},c_i)\}_{(i)}$ of pairs  associated with $\TX^{\ell}[L]$, where $c_i$ is the number of times an account $\Acck{i}$ appears in the transactions $\TX^{\ell}[L]$ under bank feature $\fj{b}$. 
	    \end{enumerate}
	     
	    \item Each bank $B_j$ partitions its account identifiers as per flag values, i.e., for every possible flag value $\flag$, it computes set $\Acc^{\flag}_{B_j}$ as the set of accounts at $B_j$ that have the flag value equal to $\flag$.
	    
	    \item PNS---with inputs $\{(\mathsf{tests}_{f_b}[L], \Acc^{0}[L], \Acc^{1}[L]) : L\in\mathcal{L}_B\}$, and banks---with inputs $\{\Acc^{\flag}_{B_j}: \flag\in\textrm{domain of flags}\}$ at bank $B_j$, run Private Intersection Sum (PI-Sum) functionality under Homomorphic Encryption (HE) (Fig.~\ref{fig:PI-sum}).
	    \begin{itemize}
	        \item As the output of PI-Sum, each bank obtains $\en{N^{\ell}_{B_j}{[L]}}$, where $N^{\ell}_{B_j}{[L]}$ denotes the number of label-$\ell$ transactions with accounts from $B_j$ that satisfy all the tests in $\tests{L}$. 
	    \end{itemize}
	    
	    \item Each bank $B_j$ applies an $\epsilon$-DP mechanism to
	    $\left\{\left(\en{N^{0}_{B_j}{[L]}},\en{N^{1}_{B_j}{[L]}}\right) \right\}_{L\in\mathcal{L}_B}$
	    (details in Sec.~\ref{sec:privacy}).
	    
	    \item Each bank $B_j$ sends the list of $\epsilon$-DP counts $\left\{\left(\en{N^{0}_{B_j}{[L]}},\en{N^{1}_{B_j}{[L]}}\right) \right\}_{L\in\mathcal{L}_B}$ to Aggregator.
	    
	    \item Aggregator computes summations under HE as follows to obtain $|\mathcal{L}_B|$ ciphertexts: $\en{\NlL{\ell}{L}} \leftarrow \sum_{j=1}^{N}\en{N^{\ell}_{B_j}{[L]}}$, and sends these ciphertexts to every bank
	    
	    \item Each bank computes a comparison under HE for each leaf node to obtain its encrypted label as: $\Label{L}:= \en{\NlL{1}{L} \stackrel{?}{<} \NlL{0}{L}} \leftarrow \compare{\en{\NlL{0}{L}}}{\en{\NlL{1}{L}}}$.
	\end{enumerate}
 	\end{minipage}
 	}
\end{figure}

\vspace{4pt}
\noindent\textbf{Private Leaf Label Computation (Fig.~\ref{fig:get-private-labels}):} To understand Step 1, consider an example of $L$ with $\tests{L} = $ $\{$Currency is USD, Ordering Account Flag is `00', Settlement Amount less than 500,000$\}$.
In this case, the associated bank feature is $\fj{b}[L] = $ Ordering Account Flag and its flag value is $\mathsf{tests}_{\fj{b}}[L]$ = {Ordering Account Flag is `00'.}
Set $\TX^{\ell}[L]$ consists of label-$\ell$ transactions that are in USD and have Settlement Amount less than 500,000. Finally, the set of candidate accounts and their corresponding counts is denoted as $\Acc^{\ell}[L] = \{(\Acck{i}, c_i)\}_{(i)}$.
This set contains account identifiers $\Acck{i}$'s of Ordering Accounts in $\TX^{\ell}[L]$, and $c_i$ is the number of transactions in which $\Acck{i}$ appears as an Ordering Account. 
In Step 2, each bank partitions its account identifiers based on their flag values.

An important component of computing labels is to obtain (under HE) the number $N^{\ell}_{B_j}[L]$ of label-$\ell$ transactions with accounts from a bank $B_j$ that satisfy all the tests in $\tests{L}$. For our example $L$, $N^{\ell}_{B_j}[L]$ is the number of label-$\ell$ transactions that have an Ordering Account from $B_j$ with Flag `00', Currency in USD and Settlement Amount less than 500,000. This is performed by running the private intersection sum (PI-Sum) functionality between PNS and each bank. Fig.~\ref{fig:PI-sum} presents the details of \textsc{PI-Sum}. At the end of \textsc{PI-Sum}, banks obtain $\en{N^{\ell}_{B_j}[L]}$. Each bank then adds a Laplace noise for DP (Sec.~\ref{sec:privacy}), and sends these noisy encrypted counts to the aggregator, which computes the summation under HE to obtain $\en{\NlL{\ell}{L}}$, the number of label-$\ell$ transactions that reach $L$. Aggregator sends these encrypted counts to all banks. 
The last step is to perform a comparison under HE at each bank to obtain an encrypted label. For efficiently computing a comparison under HE, we leverage exact techniques as in \cite{Iliashenko2021} when using schemes such as BGV \cite{bgv} or BFV \cite{bfv1, bfv2}, or leverage approximate comparisons for approximate HE schemes such as CKKS \cite{CKKS2017} as in \cite{cheon2020efficient}, where we clean up the resulting indicator masks to get higher precision by using \cite{bleach}.

\vspace{4pt}
\noindent\textbf{Private Intersection Sum (\textsc{PI-Sum}) (Fig.~\ref{fig:PI-sum}):} 
\textsc{PI-Sum} runs separately between PNS and each bank $B_j$. For each leaf $L$ and label $\ell$, PNS has the set $\Acc^{\ell}[L]$, which contains account identifiers belonging to candidate transactions that can reach $L$. Bank $B_j$ has the sets $\Acc^{\flag}_{B_j}$ of accounts that have flag value $\flag$. \textsc{PI-Sum} performs comparison under HE of the account identifiers of $\Acc^{\ell}[L]$ with the accounts in $\Acc^{\mathsf{tests}_{f_b}[L]}_{B_j}$, and sums the corresponding counts whenever they match (under HE), i.e., it computes $\en{\sum_{\{i:\Acck{i}\in\Acc^{\flag}_{B_j}\}}c_i}$. It is not difficult to prove that $ \sum_{\{i:\Acck{i}\in\Acc^{\flag}_{B_j}\}}c_i$ is equal to $N^{\ell}_{B_j}[L]$, which is the number of label-$\ell$ transactions with accounts from a bank $B_j$ that satisfy all the tests in $\tests{L}$. 
To perform this computation efficiently, we leverage  Cuckoo hashing~\cite{pagh2001_cuckoo} at PNS side, utilizing similar ideas as in~\cite{pinkas2014_faster,pinkas2019_phasing}. We realize private equality tests of $m$ bits numbers $a=(a_i)_i$, $b=(b_i)_i$, $0 \le i < m$ by $\PET{a}{b}$ under HE as a binary circuit $\left(\prod_i (1 - a_i - b_i)\right)^2$.

\begin{figure}[!h]
\small
\caption{
	Private Intersection-Sum (\textsc{PI-Sum}) for generating encrypted counts at banks.}
	\label{fig:PI-sum}
	\vspace{-0.4cm}
\fbox{
	\begin{minipage}{0.97\textwidth}
	\myparagraph{Participants:} PNS $S$, Bank $B_j$
	
	\myparagraph{Public Parameters:} a security parameter $\lambda$, an HE scheme $\E_S$, hash functions for Cuckoo hashing
	
	\myparagraph{Inputs at PNS:} Sets of candidate account identifiers with counts and associated flag values for `red' leaf nodes $\left\{(\mathsf{tests}_{f_b}[L], \Acc^{0}[L], \Acc^{1}[L])\right\}_{L\in\mathcal{L}_B}$, an HE key-pair $(\sk,\pk)$
	
	\myparagraph{Inputs at Banks:} Sets of account identifiers $\{\Acc_{B_j}^{\flag}: \flag\in\textrm{ domain of flags}\}$, an HE public-key $\pk$ 
	
    \myparagraph{Output:} Bank obtains encrypted counts for the `red' leaf nodes: $\en{N^{\ell}_{B_j}[L]}$, where $N^{\ell}_{B_j}[L]$ denotes the number of label-$\ell$ transactions with accounts from $B_j$ that reach the leaf node $L$ (i.e., pass $\mathsf{tests}_{f_b}[L]$)

	\begin{enumerate}
	   \item PNS uses Cuckoo hashing to insert, for each `red' leaf node $L\in\mathcal{L}_B$ and label $\ell\in\{0,1\}$, elements of $\Acc^{\ell}[L]$ into a hash table $\hashTable^{\ell}_L$ with $(\numBins\cdot h)$ bins using $h$ hash functions $\{\mathsf{hash}_h\}_h$. A bin of $\hashTable^{\ell}_L$ contains $(\Acck{i}, c_i)$ (or it may be empty).
	   
	   \item PNS pads all the hash tables with (dummy account no., zero count) so that all bins are occupied.
	   
	   \item PNS encrypts all the hash table element-wise. A bin of $\en{\hashTable^{\ell}_L}$ contains $(\en{\Acck{i}}, \en{c_i})$ 
	   
	   \item PNS sends $\{(\mathsf{tests}_{f_b}[L], \en{\hashTable^{0}_L}, \en{\hashTable^{1}_L})\}_{L\in\mathcal{L}_B}$ to bank $B_j$
	   
	    \item Bank $B_j$ performs private equality tests, under HE, to compare its account identifiers in $\Acc^{\flag}_{B_j}$ with entries in $\en{\hashTable^{0}_L}, \en{\hashTable^{1}_L})$ to obtain $\en{N^{\ell}_{B_j}{[L]}} \leftarrow \en{\sum_{\{i:\Acck{i}\in\Acc^{\flag}_{B_j}\}}c_i}$ for each $L, \ell$.
	    \begin{itemize}
	        \item Specifically, $B_j$ performs: $\sum_{\substack{i:\Acck{i}\in\Acc^{\mathsf{tests}_{f_b}[L]}_{B_j}\\ h\in[h]}} \PET{\en{\hashTable^{\ell}_L[\mathsf{hash}_h(\Acck{i})][1]}}{\Acck{i}}\cdot \en{\hashTable^{\ell}_L[\mathsf{hash}_h(\Acck{i})][2]}$
	    \end{itemize}
	\end{enumerate}
 	\end{minipage}
 	}
\end{figure}

\vspace{4pt}
\noindent\textbf{Resulting Model after Training:} At the end of our private training protocol~\ref{fig:priv-RDT-train},  PNS has the random forest $\mathcal{F}$ with tree structures. PNS holds the labels for `green' leaf nodes which involve only PNS features. For `red' leaf nodes, which involve PNS and bank features, every bank possesses their labels under HE encryption for which PNS holds the secret-key. This ensures that PNS does not learn any information about account datasets at the banks, and the banks do not learn any information about the PNS dataset and the labels.

\subsection{Privacy-Preserving Inference}
\label{sec:priv-RDT-inference}
Let us consider the case when PNS gets an unlabeled transaction $\tx$, and wants to predict its label. Fig.~\ref{fig:priv-RDT-inference} describes the protocol for inference. At a high-level, the algorithm computes the label for every tree in the forest and then takes the majority vote. The confidence is the fraction of the count of the majority class divided by the number of trees. 

To obtain a label for a tree $T\in\mathcal{F}$. PNS traverses $\tx$ through $T$, starting with the root node and following the arc according to which test passes. 
IF $\tx$ reaches a `green' leaf node, PNS obtains the label of the tree $T$ and continues to the next tree. Otherwise, $\tx$ reaches an internal node in $T$ associated with a bank feature $\fj{b}$. In this case, PNS cannot continue the tree traversal because it does not know the value of the bank feature in $\tx$. Then, PNS sends encrypted account number and candidate leaf nodes to banks, and the banks perform private equality tests to obtain the correct label. The details are given in Fig.~\ref{fig:priv-RDT-inference}.

\begin{figure}[!h]
\small
	\caption{
	Protocol for Privacy-Preserving Inference on Random Forest.}
	\label{fig:priv-RDT-inference}
	\vspace{-0.4cm}
\fbox{
	\begin{minipage}{0.97\textwidth}
	\myparagraph{Participants:} PNS $S$, Banks $B_1, \ldots B_N$
	
	\myparagraph{Public Parameters:} a security parameter $\lambda$, an HE scheme $\E_S$
	
	\myparagraph{Inputs at PNS:} A transaction $\tx$ for prediction, Random Forest $\mathcal{F}$, Labels for `green' leaf nodes
	
	\myparagraph{Inputs at Banks:} Encrypted labels for `red' leaf nodes
	
    \myparagraph{Output:} Label for $\tx$ along with a confidence score

	\begin{enumerate}
	    \item For each tree $T$ in the random forest $\mathcal{F}$, PNS filters out $\tx$ through $T$
	    \item If $\tx$ reaches a `green' leaf node, PNS obtains its label, and goes to the next tree
	    \item Else, $\tx$ encounters a node with bank feature $\fj{b}$ and it going to reach one of the `red' leaf nodes 
	    \item PNS computes the set $\C{\tx}$ candidate leaf nodes for $\tx$. These are the nodes $L$ in $T$ such that $\tx$ satisfies all the tests in $\tests{L}$ except the test associated with $\fj{b}$
	    \item PNS retrieves the account $\Acck{i^*}\in\tx$ corresponding to $\fj{b}$
	    \item PNS sends $\en{\Acck{i^*}}$ along with the indices of the nodes in $\C{\tx} = \{L_{k_1}, \dots, L_{k_{m'}}\}$ and their corresponding flag values $\{\mathsf{tests}_{f_b}[L_{k_p}]\}_p$ to the bank that contains $\Acck{i^*}$. (If PNS does not know which bank to send to, then it sends to all banks.)
	    \item For each $L_{k_p}$, the bank compares $\en{\Acck{i^*}}$ with its accounts with flag $\mathsf{tests}_{f_b}[L_{k_p}]$ and multiplies the result with $\en{\Label{L_{k_p}}}$.
	    \begin{itemize}
	        \item Specifically, the bank performs: $\left(\sum_{\{\Acck{i}\in\Acc^{\mathsf{tests}_{f_b}[L_{k_p}]}_{B_{j^*}}\}} \PET{\en{\Acck{i^*}}}{\Acck{i}}\right)\cdot\en{\Label{L_{k_p}}}$
	        \item It is straightforward to prove that the above computation yields $\en{\Label{L_{k_*}}}$ -- the encrypted label of the leaf node $L_{k_*}$ where $\tx$ will reach.
	    \end{itemize}
	     
	   \item Bank sends $\en{\Label{L_{k_*}}}$ to PNS, which it decrypts and goes to the next tree.
	   
	   \item After repeating the above steps for all trees, PNS outputs the label with the majority vote with confidence equal to the fraction of the majority.
	\end{enumerate}
 	\end{minipage}
 	}
\end{figure}


\section{Privacy Guarantees}
\label{sec:privacy}

\noindent\textbf{Simulation-based Privacy for Training and Inference:} We prove privacy for our training and inference protocols together in the honest-but-curios model. Recall that $\Tdb$ denotes the PNS dataset, $\Adb{i}$ denotes the account dataset at bank $B_i$, $\FS, \FB$ denote PNS and bank features respectively. Let $\mathcal{F}$ denote the random variable (r.v.) for tree structures in the random forest, and $\mathcal{L}_S$ and $\mathcal{L}_B$ denote the r.v. for sets of `green' and `red' leaf nodes in $\mathcal{F}$.
Let $R$ denote the random seed for generating $\mathcal{F}$.
Let $\{\Label{L}\}_{L\in\mathcal{L}_S}$ denote the labels of `green' leaf nodes, and $\{\Label{L}'\}_{L\in\mathcal{L}_B}$ denote the \textit{noisy} labels of `red' leaf nodes. 
We consider the \textit{view} of a participant (either PNS, bank, or aggregator) as their internal state (including their dataset, inputs, and randomness) and all messages they receive (the sent messages can be determined using the view). At a high-level, we prove privacy of sensitive datasets by giving a \textit{probabilistic polynomial-time} (PPT) simulator that can indistinguishably simulate the view of each honest party without the sensitive dataset of any other party. Intuitively, this will show that each party learns nothing about the sensitive dataset of the other parties. 
Let $\Real{\lambda}{S}{\Tdb, \FS, \FB, \mathcal{F}}$, $\Real{\lambda}{B_i}{\Tdb, \FS, \FB, \mathcal{F}}$, $\Real{\lambda}{A}{\Tdb, \FS, \FB, \mathcal{F}}$ be the random variables representing the views of PNS, Bank $B_i$, and Aggregator, respectively. 
\vspace{-4pt}
\begin{theorem}
\label{thm:privacy}
\begin{itemize}
    \item[(1)] 
    \textbf{Privacy against PNS:} There exists a PPT simulator $\mathsf{SIM}_{S}$ such that for all $\Tdb$, $\{\Adb{i}\}_{i=1}^{N}$, $\FS$, $\FB$, $R$,  
    $$\Real{\lambda}{S}{\Tdb, \FS, \FB, R, \{\Adb{i}\}_{i=1}^{N}} \approx \Sim{\lambda}{S}{\Tdb, \FS, \FB, R, \{\Label{L}\}_{L\in\mathcal{L}_S}, \{\Label{L}'\}_{L\in\mathcal{L}_B}}.$$
    
    \item[(2)] 
    \textbf{Privacy against Aggregator:} There exists a PPT simulator $\mathsf{SIM}_{A}$ such that for all $\Tdb$, $\{\Adb{i}\}_{i=1}^{N}$, $\FS$, $\FB$, $R$, 
     $\Real{\lambda}{S}{\Tdb, \FS, \FB, R, \{\Adb{i}\}_{i=1}^{N}} \approx \Sim{\lambda}{A}{|\mathcal{L}_B|}.$ 
    
    \item[(3)] 
    \textbf{Privacy against Banks:} There exists a PPT simulator $\mathsf{SIM}_{B_i}$ such that for all $\Tdb$, $\{\Adb{i}\}_{i=1}^{N}$, $\FS$, $\FB$, $R$, 
     $\Real{\lambda}{S}{\Tdb, \FS, \FB, R, \{\Adb{i}\}_{i=1}^{N}} \approx \Sim{\lambda}{R}{\Adb{i}, \FB, \mathcal{L}_B, \{\mathsf{tests}_{f_b}[L]\}_{L\in\mathcal{L}_B}}.$ 
\end{itemize}
\end{theorem}

The above theorem shows that PNS's view can be simulated given only its dataset and inputs, along with the labels. It does not learn anything about the sensitive account dataset of any bank. 
Aggregator's view can be simulated given only the number of `red' leaf nodes. It does not learn anything about the sensitive datasets of PNS and the banks.
The bank $i$'s view can be simulated given only its dataset and inputs, along with the set of `red' leaf nodes and associated flag values. Since trees are generated independently of the datasets, Bank $B_i$ does not learn anything about the sensitive datasets of PNS and the other banks.
These results can be generalized to show that even if banks collude with the aggregator, the colluding parties do not learn anything about PNS's  transaction dataset. The proofs use a standard hybrid argument, and follow from the privacy of HE and the fact that encrypted hash tables are identical in size due to padding.

\vspace{4pt}
\noindent\textbf{Differential Privacy for Protecting against Inference-Time Attacks:} We develop a DP mechanism~\cite{dwork2014algorithmic} to provide output privacy during inference. 
Since PNS obtains the labels during inference, we consider an individual account identifier as the \textit{privacy unit}. In this case, \textit{neighboring datasets} at a bank are two account datasets that differ in a single account number. 
For simplicity, let us focus on a single RDT. Due to the post-processing property of DP, it suffices to develop a DP mechanism for the counts of label-$\ell$ transactions at a leaf to protect the labels. The key step is to find the global sensitivity of the counts. 
The fact that we want to protect individual accounts makes it challenging. 
An account can participate in multiple transactions, and in the worst-case, one account could appear in all transactions. This makes the global sensitivity equal to the number of transactions, requiring a significant amount of noise and destroying the utility of the trained model. 

We enable a trade-off between privacy and accuracy by reducing the global sensitivity with a bounding technique, which has been shown to be effective in~\cite{aktay2020_google}. We use a pre-determined positive integer $\mathsf{bound}$ that is public to PNS and banks. If an account appears in more than $\mathsf{bound}$ transactions, then PNS randomly selects $\mathsf{bound}$ of them, and discards the others. In this case, it is straightforward to show that the global sensitivity of the count at a leaf node is $\mathsf{bound}$. In Step 4 in \textsc{GetPrivateLabels} (Fig.~\ref{fig:get-private-labels}), each bank applies a Laplace mechanism $\mathcal{M}$ to its encrypted input $\en{\NlL{\ell}{B_j}[L]}$ as follows
\vspace{-4pt}
\begin{equation*}
    \label{eq:DP-mechanism}
    \mathcal{M}\left(\en{\NlL{\ell}{B_j}[L]}\right) := \en{\NlL{\ell}{B_j}[L]} + \textrm{Lap}(0, \mathsf{bound}/\epsilon); \:\: \textrm{Lap}(0,\lambda): \textrm{ Laplace distribution}.
    \vspace{-2pt}
\end{equation*}
Due to the homomorphic property of the encryption, this is equivalent to directly adding the Laplace noise the count $\NlL{\ell}{B_j}[L]$ and encrypting the noisy count. Then, using standard DP arguments and leveraging the fact that tree structure does not depend on any data, we can prove that the above mechanism $\mathcal{M}$ is $\epsilon$-DP for a single RDT. 
Next, equipped with the DP mechanism for a single RDT, we want to  manage the privacy budget for training a random forest. Hence, we propose to train each tree on a randomly sampled disjoint set of transactions. This enables us to leverage the Parallel Composition Theorem of DP to prove that the entire random forest is $\epsilon$-DP. 

\section{Implementation: Security, Privacy and Scalability}

In this section, we discuss the implementation of our proposed system and its properties in terms of its security, privacy, and scalability. To implement cryptographic primitives, we use CKKS scheme~\cite{CKKS2017} using the HEaaN library \cite{heaan}, implementing low-level features in C and non-critical functions in Python. 

\subsection{Security of Cryptographic Primitives}
In our prototype, we assume that Flower provides secure communication channels preventing man-in-the-middle attacks. We assume that all the keys that are involved in the protocols are stored securely in some Key Management Device (KMS). Currently, our implementation is a stand-alone implementation and thus does not include a KMS. However, we emphasize that special care is needed when handling cryptographic keys.

Our system relies on HE and more specifically on CKKS cryptographic primitives.
We now present potential attacks on these systems.
 When dealing with cryptographic primitives instead of specifying and evaluating a specific attack we describe a property that basically states what a probable polynomial time (PPT) adversary can expect to achieve no matter how it facilitates the attack.

    \noindent\textbf{CKKS is semantic security and IND-CPA.} For a given security parameter, every PPT adversary $\mathcal{A}$ who has access to an encryption and evaluation oracles (who can use the public and evaluation keys, respectively), can generate $n$ messages $m_1, \ldots, m_n$ and their encryptions $Enc(m_1), \ldots, Enc(m_n)$, respectively. However, given encryption $Enc(m_0)$ of an additional message $m_0$ unknown to $\mathcal{A}$, it cannot guess $m_0$ with a probability that is non-negligible in the security parameter. Hence, for potential adversaries with access to ciphertext or public keys, no information is revealed. 
    We use the HEaaN library \cite{heaan} for CKKS implementation configured with security parameter $\lambda=128$, and we assume that it fulfills the above property.

    

    \noindent \textbf{Circuit Privacy.} An HE is considered circuit private if for every PPT adversary $\mathcal{A}$ who holds the HE keys (secret key, public key, and evaluation keys), given a ciphertext $\Enc{m} = C(inputs)$ which is the output of a secret circuit, $C$, computed under FHE, $\mathcal{A}$ on secret $inputs$ cannot learn anything about $C$ nor its inputs. 
    Specifically, FHE schemes such as BGV and CKKS are not circuit private. In fact, some works show that the noise encoded in a ciphertext does leak info on the circuit or its inputs. While some works show that it may be feasible to extract data from decryption of ciphertexts, we are unaware of any robust attack that can do so for complex circuits. If one such attack is demonstrated, a potential mitigation called \textit{sanitization} was presented in~\cite{ducas2016sanitization}, where the noise is injected in the ciphertext resulting in no information on $C$ leaks from $Enc(m)$ with overwhelming probability.

    In the case of circuit privacy, we distinguish between three types of threats: Extracting the $inputs$ (of the banks) by PNS, extracting the circuit $C$ of the banks, e.g., the number of accounts, or learn from approximation errors e.g., during comparisons.

    \textbf{IND-CPA$^{+}$.} Li and Micciancio~\cite{li2021on} showed an attack on IND-CPA CKKS where an adversary who holds an encryption of a message $m$ - $Enc(m)$ and its decryption may be able to extract the secret key. Many cryptographic libraries already include some mitigation in the form of injecting noise before performing the decryption. We consider the implementation based on the HEAAN library and inherit its mitigation if available. Alternatively,
    if the PNS does not trust the underlying library it can round the decryption output (label) to the closest integer, which also mitigates the above issue.

\subsection{Scalability}
\label{sec:scalability}

The key cryptographic component in our solution is the HE cryptosystem. 
Modern HE instantiations such as BGV \cite{bgv}, B/FV \cite{bfv1, bfv2}, and CKKS~\cite{CKKS2017} can operate on ciphertexts in a homomorphic single instruction multiple data (SIMD) fashion. This means that a single ciphertext encrypts a fixed-size vector, and the homomorphic operations on the ciphertext are performed slot-wise on the elements of the plaintext vector. Our protocols are highly parallelizable in terms of HE operations (e.g., hash table comparison for different leaf nodes can be parallelized), and in addition, leverage the SIMD capabilities to achieve efficiency at scale. 


Cryptographic operations take place only for branches with red leaves.
Thus, to understand scalability, in the following, we analyze the effect of the expected number of red leaves.
We first restate the process in which we pick the condition checked in a node of the tree.
We start drawing conditions from the root and move down toward the leaves in a breadth-first manner.
Denote by $\CS$ the set of conditions that involve PNS features and by $\CB$ the set of conditions that involve the Bank's features.
At a node $v$, we draw a condition in the following way: if $v$ has an ancestor for whom we drew a condition from $\CB$, we draw a random condition from
$\CS$ otherwise, we draw a condition from $\CB \cup \CS$.

Denote by $R_i$ the random variable counting the red leaf nodes in an RDT with $i$ levels of inner nodes. The following lemma gives the expected value or $R_i$.
\begin{lemma}
\label{lem:number-of-red-leaves}
The expected number of red leaf nodes in an RDT with $i$ levels of inner nodes is given as $\mathbb{E}(R_{i}) = 2^i (1- \frac{|\CS|^i}{(|\CS|+|\CB|)^i}).$
\end{lemma}

\begin{proof}
We have
$$
R_1 = \begin{cases}
    0, & \text{ with probability } \frac{|\CS|}{|\CS|+|\CB|}\\
    2, & \text{ with probability } \frac{|\CB|}{|\CS|+|\CB|}
\end{cases}
\quad\text{ and }\quad
R_{i+1} = \begin{cases}
    2R_i, & \text{ with probability } \frac{|\CS|}{|\CS|+|\CB|}\\
    2^{i+1}, & \text{ with probability } \frac{|\CB|}{|\CS|+|\CB|}
\end{cases}
.$$
It is easy to see that $\mathbb{E}(R_1) = 2\frac{|\CB|}{|\CS|+|\CB|}$ and
$\mathbb{E}(R_{i+1}) = 2^{i+1}\frac{|\CB|}{|\CS|+|\CB|} + 2\mathbb{E}(R_i)\frac{|\CS|}{|\CS|+|\CB|}$. This solves to:
$$\mathbb{E}(R_{i}) = 2^i \Big(1- \frac{|\CS|^i}{(|\CS|+|\CB|)^i}\Big),$$
which proves the Lemma.
\end{proof}

Some works have shown that using a random forest with the number of RDTs comparable to the number of features and with a height less than or equal to half the number of features yields a favorable utility versus privacy trade-off when using DP~\cite{Jagannathan2012_apractical, Fletcher2017_differentially, Fletcher2019_decision}. Since the number of features in typical financial anomaly detection scenarios is small (around 10-20), we consider $\CS = 10$ PNS conditions, $\CB= 2$ bank conditions, and $N_{trees} = 12$ trees, each with height $h = 6$.

We plug the parameters we use into Lemma~\ref{lem:number-of-red-leaves} to get the expected number of red leaves is:
$$\text{Expected number of red leaves} = N \cdot \mathbb{E}(R_{h}) = N \cdot 2^h \Big(1- \frac{|\CS|^h}{(|\CS|+|\CB|)^h}\Big) = 12 \cdot 2^6 \Big(1- \frac{10^6}{(10+2)^6}\Big) \approx 511.$$

In our implementation, we consider 800 read leaf nodes to get a good estimate of scalability and efficiency.

\subsection{Experimentally benchmarking cryptographic flows}
\label{sec:experiments}
For all the experiments, we considered an Intel\circler~ Xeon\circler~ CPU E5-2699 v4 @ 2.60GHz with 16 cores and Tesla P100-PCIE-16GB GPU with 124GB memory. 
Note that the dominant part of the computation was done on the GPU leaving the CPU mostly idle. In practice, PNS only encrypts and decrypts messages which is less computation intensive than the computation done by the banks.
In our experiments, we used the HEaaN library \cite{heaan} for the implementation of the CKKS scheme~\cite{CKKS2017}. We used preset parameters where ciphertexts have $2^{15}$ slots, a multiplication depth of $9$, fractional part precision of $24$, and integer part precision of $5$. This context allows us to use up to $6$ multiplications before bootstrapping is required. 

\begin{table}[]
    \centering
    \caption{Performance benchmarks for the cryptographic parts, where the number of banks is 1 and the number of leaves is 800. ($UA_{PNS}$) - number of PNS unique accounts, ($UA_{Bank}$) - number of banks unique accounts, $t_{inf}$ number of transaction during  inference. Here, 'k' is used to denote 1000, in other words, 100k denotes 100000, 1500k denotes 1500000, and so on.}
    \label{tab:exphe}
    \begin{tabular}{|c|c|c|c|c|c|}
        \hline
        Red &$UA_{PNS}$ & $UA_{Bank}$ & $t_{inf}$ & Training time & Inference time \\
        leaves & &  &  & (min) & (min) \\
        \hline
        800 & 100k & 100k  & 1500k & 16  & 130 \\
        800 & 100k & 100k & 1000k & 18  & 53  \\
        800 & 200k & 100k & 1000k & 35  & 54  \\
        800 & 300k & 100k & 1000k & 51 & 55\\
        800 & 400k & 100k & 1000k& 60 & 55\\
        800 & 100k & 500k & 1000k & 20 & 58 \\
        800 & 100k & 1500k & 1000k & 21 & 55\\
        800 & 100k & 2000k & 1000k & 21 & 53\\
        \hline
    \end{tabular}
\end{table}

Table~\ref{tab:exphe} presents some of our experiments. As discussed in Sec.~\ref{sec:scalability}, we consider 800 read leaf nodes to get a good estimate of scalability
and efficiency. 
We fixed the number of banks to 1 since running the prediction protocol involves PNS and 1 bank. We emphasize that PNS runs multiple protocol instances with multiple banks in parallel, but in each instance, there is only 1 bank participating.
We also set the maximal number of transactions per unique account during inference to 15.

The experiments we made simulated only the secure computation part. Specifically, it did not include the machine learning parts (which happen in plaintext) or the differential privacy parts. The results show that our solution is feasible, but also practical even when using a commodity GPU. Moreover, a careful look into the details shows that every computation is made of sequential batches.

{\bf Training time:}
The training time depends on the number of \textit{unique} accounts PNS has ($UA_{PNS}$) because multiple transactions of the same account are joined together.
Specifically, PNS generates a hash table that is linear in $UA_{PNS}$.
Then, each bank compares its set of unique accounts ($UA_{Bank}$) to that hash table. The training time is therefore proportional to the product of $UA_{PNS} \times UA_{Bank}$ and the number of leaves (which is fixed to 800 in our experiments).
Indeed, our experiments corroborate this. For example, when fixing $UA_{Bank}=100k$ we have a training time of 18, 35, 51, and 60 minutes when $UA_{PNS}=100000, 200000, 300000,$ and $400000$ (resp.). When fixing $UA_{PNS} = 100000$ we have training time of 18,20 and 21 minutes when $UA_{Bank}=100000, 500000,$ and $1500000$ (resp.). This is because there is a dominant constant overhead time and a small linear dependency on $UA_{Bank}$.

{\bf Inference time:}
We see that the inference time is scalable and can support large values for $UA_{Bank}$ and $UA_{PNS}$ without changing the inference time.
Since labeling transactions is embarrassingly parallelizable, it is easy to support more transactions by adding more hardware and by splitting the set of transactions into smaller batches. For example, instead of performing 1,500,000 transactions (Table \ref{tab:exphe} First line), it is possible to achieve better results by running batches of 1,000,000 transactions.

Overall, our experiments demonstrate that our solution is feasible, scalable, and can support systems with a large number of accounts and transactions to label. 
\section{Conclusion}
\label{sec:conclusion}
We have presented, PV4FAD, a solution for privacy preserving training and inference of a predictor for financial anomaly detection that addresses the vertical and horizontal partitioning of data to train a vertical ensemble model. 
The core of our solution is the use of a random decision tree that the PNS and banks train collaboratively. Tree node labels can be obtained by PNS for nodes that only involve features that PNS has. Banks obtain encrypted leaf node labels using a homomorphic encryption-based protocol when features from both banks and PNS are needed. We also use differential privacy to prevent inference threats and inhibit the transfer of data through cryptographic primitives.  At inference time, banks and PNS use a homomorphic encryption-based protocol for PNS to generate predictions.

This approach provides provable privacy guarantees for input privacy as well as differential privacy guarantees for output privacy against inference attacks with relatively low overhead. It also enables the use of standard and complex features as needed for a complex financial anomaly detection system. 
This approach is generalizable to other scenarios with asymmetric data ownership patterns.

\bibliographystyle{plain} 
\bibliography{references} 

\begin{thebibliography}{10}

\bibitem{weforum}
{Financial Crime by World Economic Forum}.
\newblock
  \url{https://www.weforum.org/agenda/2018/01/we-need-to-talk-about-financial-crime/}.

\bibitem{unitednations}
{Money Laundering Overview by United Nations}.
\newblock \url{https://www.unodc.org/unodc/en/money-laundering/overview.html}.

\bibitem{pprl2022}
Allon Adir, Ehud Aharoni, Nir Drucker, Eyal Kushnir, Ramy Masalha, Michael
  Mirkin, and Omri Soceanu.
\newblock Privacy-preserving record linkage using local sensitive hash and
  private set intersection.
\newblock In {\em To appear in: Applied Cryptography and Network Security
  Workshops}, Cham, 2022. Springer International Publishing.

\bibitem{aktay2020_google}
Ahmet Aktay, Shailesh Bavadekar, Gwen Cossoul, John Davis, Damien Desfontaines,
  Alex Fabrikant, Evgeniy Gabrilovich, Krishna Gadepalli, Bryant Gipson, Miguel
  Guevara, Chaitanya Kamath, Mansi Kansal, Ali Lange, Chinmoy Mandayam, Andrew
  Oplinger, Christopher Pluntke, Thomas Roessler, Arran Schlosberg, Tomer
  Shekel, Swapnil Vispute, Mia Vu, Gregory Wellenius, Brian Williams, and
  Royce~J Wilson.
\newblock Google covid-19 community mobility reports: Anonymization process
  description (version 1.1), 2020.

\bibitem{crf}
Louis J.~M. Aslett, Pedro~M. Esperan{\c{c}}a, and Chris~C. Holmes.
\newblock {Encrypted statistical machine learning: new privacy preserving
  methods}.
\newblock pages 1--21, 2015.

\bibitem{bonawits2017_practical}
Keith Bonawitz, Vladimir Ivanov, Ben Kreuter, Antonio Marcedone, H.~Brendan
  McMahan, Sarvar Patel, Daniel Ramage, Aaron Segal, and Karn Seth.
\newblock Practical secure aggregation for privacy-preserving machine learning.
\newblock In {\em Proceedings of the 2017 ACM SIGSAC Conference on Computer and
  Communications Security}, CCS'17, page 1175–1191, 2017.

\bibitem{bfv2}
Zvika Brakerski.
\newblock {Fully Homomorphic Encryption without Modulus Switching from
  Classical GapSVP}.
\newblock In Reihaneh Safavi-Naini and Ran Canetti, editors, {\em Advances in
  Cryptology -- CRYPTO 2012}, volume 7417 LNCS, pages 868--886, Berlin,
  Heidelberg, 2012. Springer Berlin Heidelberg.

\bibitem{bgv}
Zvika Brakerski, Craig Gentry, and Vinod Vaikuntanathan.
\newblock {(Leveled) Fully Homomorphic Encryption without Bootstrapping}.
\newblock {\em ACM Transactions on Computation Theory}, 6(3), jul 2014.

\bibitem{cheng2021secureboost}
Kewei Cheng, Tao Fan, Yilun Jin, Yang Liu, Tianjian Chen, Dimitrios
  Papadopoulos, and Qiang Yang.
\newblock Secureboost: A lossless federated learning framework.
\newblock {\em IEEE Intelligent Systems}, 36(6):87--98, 2021.

\bibitem{CKKS2017}
Jung Cheon, Andrey Kim, Miran Kim, and Yongsoo Song.
\newblock {Homomorphic Encryption for Arithmetic of Approximate Numbers}.
\newblock In {\em Proceedings of Advances in Cryptology - ASIACRYPT 2017},
  pages 409--437. Springer Cham, 11 2017.

\bibitem{cheon2020efficient}
Jung~Hee Cheon, Dongwoo Kim, and Duhyeong Kim.
\newblock Efficient homomorphic comparison methods with optimal complexity.
\newblock In {\em International Conference on the Theory and Application of
  Cryptology and Information Security}, pages 221--256. Springer, 2020.

\bibitem{choquette-choo2021_label}
Christopher~A. Choquette-Choo, Florian Tramer, Nicholas Carlini, and Nicolas
  Papernot.
\newblock Label-only membership inference attacks.
\newblock In Marina Meila and Tong Zhang, editors, {\em Proceedings of the 38th
  International Conference on Machine Learning}, volume 139 of {\em Proceedings
  of Machine Learning Research}, pages 1964--1974. PMLR, 18--24 Jul 2021.

\bibitem{heaan}
CryptoLab.
\newblock {HEaaN: Homomorphic Encryption for Arithmetic of Approximate
  Numbers}, 2022.

\bibitem{bleach}
Nir Drucker, Guy Moshkowich, Tomer Pelleg, and Hayim Shaul.
\newblock {BLEACH: Cleaning Errors in Discrete Computations over CKKS}, 2022.
\newblock Manuscript, to be published.

\bibitem{ducas2016sanitization}
L{\'e}o Ducas and Damien Stehl{\'e}.
\newblock Sanitization of fhe ciphertexts.
\newblock In Marc Fischlin and Jean-S{\'e}bastien Coron, editors, {\em Advances
  in Cryptology -- EUROCRYPT 2016}, pages 294--310, Berlin, Heidelberg, 2016.
  Springer Berlin Heidelberg.

\bibitem{dwork2014algorithmic}
Cynthia Dwork, Aaron Roth, et~al.
\newblock The algorithmic foundations of differential privacy.
\newblock {\em Foundations and Trends{\textregistered} in Theoretical Computer
  Science}, 9(3--4):211--407, 2014.

\bibitem{bfv1}
Junfeng Fan and Frederik Vercauteren.
\newblock {Somewhat Practical Fully Homomorphic Encryption}.
\newblock {\em Proceedings of the 15th international conference on Practice and
  Theory in Public Key Cryptography}, pages 1--16, 2012.

\bibitem{Fan2003_is}
W.~Fan, H.~Wang, P.S. Yu, and S.~Ma.
\newblock Is random model better? on its accuracy and efficiency.
\newblock In {\em Third IEEE International Conference on Data Mining}, pages
  51--58, 2003.

\bibitem{Fletcher2017_differentially}
Sam Fletcher and Md~Zahidul Islam.
\newblock Differentially private random decision forests using smooth
  sensitivity.
\newblock {\em Expert Syst. Appl.}, 78(C):16–31, jul 2017.

\bibitem{Fletcher2019_decision}
Sam Fletcher and Md.~Zahidul Islam.
\newblock Decision tree classification with differential privacy: A survey.
\newblock {\em ACM Comput. Surv.}, 52(4), aug 2019.

\bibitem{fredrikson2015model}
Matt Fredrikson, Somesh Jha, and Thomas Ristenpart.
\newblock Model inversion attacks that exploit confidence information and basic
  countermeasures.
\newblock In {\em Proceedings of the 22nd ACM SIGSAC Conference on Computer and
  Communications Security}, CCS '15, page 1322–1333, New York, NY, USA, 2015.
  Association for Computing Machinery.

\bibitem{Friedman2010_data}
Arik Friedman and Assaf Schuster.
\newblock Data mining with differential privacy.
\newblock In {\em Proceedings of the 16th ACM SIGKDD International Conference
  on Knowledge Discovery and Data Mining}, page 493–502, New York, NY, USA,
  2010. Association for Computing Machinery.

\bibitem{ganju2018property}
Karan Ganju, Qi~Wang, Wei Yang, Carl~A. Gunter, and Nikita Borisov.
\newblock Property inference attacks on fully connected neural networks using
  permutation invariant representations.
\newblock In {\em Proceedings of the 2018 ACM SIGSAC Conference on Computer and
  Communications Security}, CCS '18, page 619–633, New York, NY, USA, 2018.
  Association for Computing Machinery.

\bibitem{Geurts2006_extremely}
Pierre Geurts, Damien Ernst, and Louis Wehenkel.
\newblock Extremely randomized trees.
\newblock {\em Mach. Learn.}, 63(1):3–42, apr 2006.

\bibitem{hardy2017private}
Stephen Hardy, Wilko Henecka, Hamish Ivey-Law, Richard Nock, Giorgio Patrini,
  Guillaume Smith, and Brian Thorne.
\newblock Private federated learning on vertically partitioned data via entity
  resolution and additively homomorphic encryption.
\newblock {\em arXiv preprint arXiv:1711.10677}, 2017.

\bibitem{he2020fedml}
Chaoyang He, Songze Li, Jinhyun So, Xiao Zeng, Mi~Zhang, Hongyi Wang, Xiaoyang
  Wang, Praneeth Vepakomma, Abhishek Singh, Hang Qiu, et~al.
\newblock Fedml: A research library and benchmark for federated machine
  learning.
\newblock {\em arXiv preprint arXiv:2007.13518}, 2020.

\bibitem{Iliashenko2021}
Ilia Iliashenko and Vincent Zucca.
\newblock {Faster homomorphic comparison operations for BGV and BFV}.
\newblock {\em Proceedings on Privacy Enhancing Technologies},
  2021(3):246--264, 2021.

\bibitem{ion2020_on}
Mihaela Ion, Ben Kreuter, Ahmet~Erhan Nergiz, Sarvar Patel, Shobhit Saxena,
  Karn Seth, Mariana Raykova, David Shanahan, and Moti Yung.
\newblock On deploying secure computing: Private
  intersection-sum-with-cardinality.
\newblock In {\em 2020 IEEE European Symposium on Security and Privacy
  (EuroS\&P)}, pages 370--389, 2020.

\bibitem{ion2017_private}
Mihaela Ion, Ben Kreuter, Erhan Nergiz, Sarvar Patel, Shobhit Saxena, Karn
  Seth, David Shanahan, and Moti Yung.
\newblock Private intersection-sum protocol with applications to attributing
  aggregate ad conversions.
\newblock {\em {IACR} Cryptol. ePrint Arch.}, page 738, 2017.

\bibitem{Jagannathan2012_apractical}
Geetha Jagannathan, Krishnan Pillaipakkamnatt, and Rebecca~N. Wright.
\newblock A practical differentially private random decision tree classifier.
\newblock {\em Trans. Data Privacy}, 5(1):273–295, apr 2012.

\bibitem{japkowicz2000class}
Nathalie Japkowicz.
\newblock The class imbalance problem: Significance and strategies.
\newblock In {\em Proc. of the Int’l Conf. on Artificial Intelligence},
  volume~56, pages 111--117. Citeseer, 2000.

\bibitem{japkowicz2002class}
Nathalie Japkowicz and Shaju Stephen.
\newblock The class imbalance problem: A systematic study.
\newblock {\em Intelligent data analysis}, 6(5):429--449, 2002.

\bibitem{kadhe2020_fastsecagg}
Swanand Kadhe, Nived Rajaraman, O.~Ozan Koyluoglu, and Kannan Ramchandran.
\newblock Fastsecagg: Scalable secure aggregation for privacy-preserving
  federated learning.
\newblock In {\em CCS Workshop on Privacy Preserving Machine Learning in
  Practice}, CCS-PPMLP, 2020.

\bibitem{kairouz2021advances}
Peter Kairouz, H~Brendan McMahan, Brendan Avent, Aur{\'e}lien Bellet, Mehdi
  Bennis, Arjun~Nitin Bhagoji, Kallista Bonawitz, Zachary Charles, Graham
  Cormode, Rachel Cummings, et~al.
\newblock Advances and open problems in federated learning.
\newblock {\em Foundations and Trends{\textregistered} in Machine Learning},
  14(1--2):1--210, 2021.

\bibitem{li2021on}
Baiyu Li and Daniele Micciancio.
\newblock On the security of homomorphic encryption on approximate numbers.
\newblock In Anne Canteaut and Fran{\c{c}}ois-Xavier Standaert, editors, {\em
  Advances in Cryptology -- EUROCRYPT 2021}, pages 648--677, Cham, 2021.
  Springer International Publishing.

\bibitem{li2020federated}
Tian Li, Anit~Kumar Sahu, Ameet Talwalkar, and Virginia Smith.
\newblock Federated learning: Challenges, methods, and future directions.
\newblock {\em IEEE Signal Processing Magazine}, 37(3):50--60, 2020.

\bibitem{zheng2021_membership}
Zheng Li and Yang Zhang.
\newblock Membership leakage in label-only exposures.
\newblock In {\em Proceedings of the 2021 ACM SIGSAC Conference on Computer and
  Communications Security}, CCS '21, page 880–895, 2021.

\bibitem{liu2008exploratory}
Xu-Ying Liu, Jianxin Wu, and Zhi-Hua Zhou.
\newblock Exploratory undersampling for class-imbalance learning.
\newblock {\em IEEE Transactions on Systems, Man, and Cybernetics, Part B
  (Cybernetics)}, 39(2):539--550, 2008.

\bibitem{liu2021fate}
Yang Liu, Tao Fan, Tianjian Chen, Qian Xu, and Qiang Yang.
\newblock Fate: An industrial grade platform for collaborative learning with
  data protection.
\newblock {\em J. Mach. Learn. Res.}, 22(226):1--6, 2021.

\bibitem{ludwig2022federated}
Heiko Ludwig and Nathalie Baracaldo, editors.
\newblock {\em Federated Learning - {A} Comprehensive Overview of Methods and
  Applications}.
\newblock Springer, 2022.

\bibitem{ludwig2020ibm}
Heiko Ludwig, Nathalie Baracaldo, Gegi Thomas, Yi~Zhou, Ali Anwar, Shashank
  Rajamoni, Yuya Ong, Jayaram Radhakrishnan, Ashish Verma, Mathieu Sinn, et~al.
\newblock Ibm federated learning: an enterprise framework white paper v0. 1.
\newblock {\em arXiv preprint arXiv:2007.10987}, 2020.

\bibitem{molloy2016graph}
I.~Molloy, S.~Chari, U.~Finkler, M.~Wiggerman, C.~Jonker, T.~Habeck, Y.~Park,
  F.~Jordens, and R.~van Schaik.
\newblock Graph analytics for real-time scoring of cross-channel transactional
  fraud.
\newblock {\em Financial Cryptography and Data Security 2016}, 2016.

\bibitem{nasr2019comprehensive}
Milad Nasr, Reza Shokri, and Amir Houmansadr.
\newblock Comprehensive privacy analysis of deep learning: Passive and active
  white-box inference attacks against centralized and federated learning.
\newblock In {\em 2019 IEEE Symposium on Security and Privacy (SP)}, pages
  739--753, 2019.

\bibitem{pagh2001_cuckoo}
Rasmus Pagh and Flemming~Friche Rodler.
\newblock Cuckoo hashing.
\newblock In Friedhelm~Meyer auf~der Heide, editor, {\em Algorithms --- ESA
  2001}, pages 121--133, Berlin, Heidelberg, 2001. Springer Berlin Heidelberg.

\bibitem{pinkas2019_phasing}
Benny Pinkas, Thomas Schneider, Gil Segev, and Michael Zohner.
\newblock Phasing: Private set intersection using permutation-based hashing.
\newblock In {\em 24th USENIX Security Symposium (USENIX Security 15)}, pages
  515--530, Washington, D.C., August 2015. USENIX Association.

\bibitem{pinkas2019_efficient}
Benny Pinkas, Thomas Schneider, Oleksandr Tkachenko, and Avishay Yanai.
\newblock Efficient circuit-based psi with linear communication.
\newblock In Yuval Ishai and Vincent Rijmen, editors, {\em Advances in
  Cryptology -- EUROCRYPT 2019}, pages 122--153, Cham, 2019. Springer
  International Publishing.

\bibitem{pinkas2014_faster}
Benny Pinkas, Thomas Schneider, and Michael Zohner.
\newblock Faster private set intersection based on {OT} extension.
\newblock In {\em 23rd USENIX Security Symposium (USENIX Security 14)}, pages
  797--812, San Diego, CA, August 2014. USENIX Association.

\bibitem{psswc2}
Y.~Athur Raghuvir, S.~Govindarajan, S.~Vijayakumar, P.~Yadlapalli, and
  F.~Di~Troia.
\newblock Advancement on security applications of private intersection sum
  protocol.
\newblock In Kohei Arai, editor, {\em Proceedings of the Future Technologies
  Conference (FTC) 2021, Volume 3}, pages 104--116, Cham, 2022. Springer
  International Publishing.

\bibitem{Rana2015_differentially}
Santu Rana, Sunil~Kumar Gupta, and Svetha Venkatesh.
\newblock Differentially private random forest with high utility.
\newblock In {\em 2015 IEEE International Conference on Data Mining}, pages
  955--960, 2015.

\bibitem{romanini2021pyvertical}
Daniele Romanini, Adam~James Hall, Pavlos Papadopoulos, Tom Titcombe, Abbas
  Ismail, Tudor Cebere, Robert Sandmann, Robin Roehm, and Michael~A Hoeh.
\newblock Pyvertical: A vertical federated learning framework for multi-headed
  splitnn.
\newblock {\em arXiv preprint arXiv:2104.00489}, 2021.

\bibitem{shalev-shwartz2014_understanding}
Shai Shalev-Shwartz and Shai Ben-David.
\newblock {\em Understanding Machine Learning: From Theory to Algorithms}.
\newblock 2014.

\bibitem{shokri2017membership}
Reza Shokri, Marco Stronati, Congzheng Song, and Vitaly Shmatikov.
\newblock Membership inference attacks against machine learning models.
\newblock In {\em 2017 IEEE symposium on security and privacy (SP)}, pages
  3--18. IEEE, 2017.

\bibitem{so2020turbo}
Jinhyun So, Basak Guler, and Amir~Salman Avestimehr.
\newblock Turbo-aggregate: Breaking the quadratic aggregation barrier in secure
  federated learning.
\newblock {\em IEEE Journal on Selected Areas in Information Theory},
  2:479--489, 2020.

\bibitem{suzumura2022federated}
Toyotaro Suzumura, Yi~Zhou, Ryo Kawahara, Nathalie Baracaldo, and Heiko Ludwig.
\newblock Federated learning for collaborative financial crimes detection.
\newblock In {\em Federated Learning}, pages 455--466. Springer, 2022.

\bibitem{truex2019_hybrid}
Stacey Truex, Nathalie Baracaldo, Ali Anwar, Thomas Steinke, Heiko Ludwig, Rui
  Zhang, and Yi~Zhou.
\newblock A hybrid approach to privacy-preserving federated learning.
\newblock In {\em Proceedings of the 12th ACM Workshop on Artificial
  Intelligence and Security}, AISec'19, page 1–11, 2019.

\bibitem{weber2018scalable}
Mark Weber, Jie Chen, Toyotaro Suzumura, Aldo Pareja, Tengfei Ma, Hiroki
  Kanezashi, Tim Kaler, Charles~E Leiserson, and Tao~B Schardl.
\newblock Scalable graph learning for anti-money laundering: A first look.
\newblock {\em arXiv preprint arXiv:1812.00076}, 2018.

\bibitem{xu2021fedv}
Runhua Xu, Nathalie Baracaldo, Yi~Zhou, Ali Anwar, James Joshi, and Heiko
  Ludwig.
\newblock Fedv: Privacy-preserving federated learning over vertically
  partitioned data.
\newblock In {\em Proceedings of the 14th ACM Workshop on Artificial
  Intelligence and Security}, pages 181--192, 2021.

\bibitem{yang2019federated}
Qiang Yang, Yang Liu, Tianjian Chen, and Yongxin Tong.
\newblock Federated machine learning: Concept and applications.
\newblock {\em ACM Transactions on Intelligent Systems and Technology (TIST)},
  10(2):1--19, 2019.

\end{thebibliography}

\end{document}